# On Using Non-Kekulé Triangular Graphene Quantum Dots for Scavenging Hazardous Sulfur Hexafluoride Components

**Vaishali Roondhe[1,*], Basant Roondhe[2], Sumit Saxena[2], Rajeev Ahuja[3,4,*] and Alok Shukla[1,*]**

[1]Department of Physics, Indian Institute of Technology Bombay, Mumbai-400076, Maharashtra, India

[2]Department of Metallurgical Engineering and Materials Science, Indian Institute of Technology Bombay, Mumbai-400076, Maharashtra, India

[3]Materials Theory Division, Department of Physics and Astronomy, Uppsala University, Box 516, Uppsala 75120, Sweden

[4]Department of Physics, Indian Institute of Technology Ropar-140001, Punjab, India

## Abstract

The goal of present study is to explore how the size and functionalization of graphene quantum dots (GQDs) affect their sensing capabilities. Specifically, we investigated the adsorption of $SO_2$, $SOF_2$, $SO_2F_2$, and $SF_6$ on GQDs that were functionalized with $-CH_3$, $-COCH_3$, and $-NH_2$. We used density functional theory to analyse the electronic properties of these functionalized GQDs and found that the functionalization significantly altered their electronic properties. For example, the B3LYP H-L gap of pristine triangulene was 3.9eV, while the H-L gap of functionalized triangulene ranged from 2.8 eV-3.6 eV (using the B3LYP functional). Our results indicate that $-NH_2$ functionalized phenalenyl and triangulene provide strong interaction with $SO_2$, with adsorption energies of -0.429 eV and -0.427 eV, respectively. These adsorption properties exhibit physisorption, leading to high gas sensitivity and superior recovery time. The findings of this study provide new insights into the potential use of GQDs for detecting the decomposed constituents of sulfur hexafluoride, which can be beneficial for assessing the operation status of $SF_6$ insulated devices. Overall, our calculations suggest that functionalized GQDs can be employed in gas insulated systems for partial discharge detection.

---

[*] Corresponding Author:
Email addresses: shukla@phy.iitb.ac.in (A. Shukla), oshivaishali@gmail.com (V. Roondhe), rajeev.ahuja@physics.uu.se (R. Ahuja)

# 1. Introduction

Sulfur hexafluoride ($SF_6$) is an odourless, nontoxic, non-flammable, colourless and inert gas that possesses excellent insulation and strong chemical stability [1-2]. Its dielectric strength and chemical inertness make it a crucial medium for electrical insulators and in gas-insulated switchgear (GIS) for power systems as an arc-quenching medium [3-4]. However, in certain cases, partial discharge can occur when the electric field is intensified over time, leading to the decomposition of SF6. The resulting products, such as $SO_2$, $H_2S$, $SO_2F_2$, $CF_4$, and $SOF_2$, are responsible for equipment corrosion and can potentially lead to system failure [5-9]. Current methods used to detect these decomposed products of $SF_6$ include ultrasonic, ultrahigh frequency, gas sensing and optical measurement methods [10-12]. Among these, the gas sensing method is highly recommended due to its high sensitivity in less volume and the most important thing, it is cost effective [13-16]. However, gases such as $SO_2$, $SOF_2$, and $SO_2F_2$, which have extremely low concentrations on the order of parts per million (PPM), are challenging to detect with conventional sensors [17].

Nanomaterials based on graphene have received significant attention in the scientific community following the successful fabrication of graphene [18]. For both fundamental science and practical applications, graphene oxide (GO), reduced graphene oxide (rGO), and graphene quantum dots (GQD) have been extensively studied [19-22]. Within the graphene family, GQD possess superior properties such as chemical inertness, low cytotoxicity, size and shape dependent photoluminescence etc. [23-25], make them useful for various applications in electronics (super capacitor, flash memory etc.), optical (photodetector, phototransistor etc.), medical (drug delivery, cancer phototherapy etc.) and energy [26-32]. Due to large surface-to-volume ratio, GQD also have superior sensing properties [24,33] making them useful in electrical and optical gas sensing devices [33-34]. A group of GQD have closed-shell electronic structures, while some, due to their unique topology, may have open-shell structure with high-spin ground states [35-36]. Open-shell GQD with π-electrons near the Fermi level have a delocalized radical character, which is useful for spintronics [37-39] and energy-related applications [40]. Closed-shell graphene fragments have closed-shell electronic configuration (no unpaired electrons) with π-electrons in bonding orbitals. However, open-shell graphene fragments have unpaired electrons or partially unpaired electrons, which results in non-bonding single occupied molecular orbitals (SOMO) in molecular orbital theory. For example, graphene fragments with two unpaired electrons are also known as biradical hydrocarbons, in which the ground state has two nonbonding molecular orbitals filled by two unpaired electrons [41]. Open-shell graphene fragments are unique in nature, possessing intriguing optical, electronic, magnetic properties and crystalline structures as compared to their closed-shell counterparts. As a result, open-shell graphene fragments have numerous potential applications in photovoltaic devices [42], spintronic devices [43-44] etc. As stated by Lambert, "The future of these biradical PAHs clearly lies in materials science" [45]. Although useful in their pristine form, oxygen (−OH, −O−, −OCH$_3$, −COOH etc), amine (−NH$_2$) and methyl (−CH$_3$) functionalized GQDs generally have improved solubility and superior quantum yield (QY) due to changes in their electronic structure caused by functionalization [46-48].

The study of sensing SF$_6$ decomposition products on GQDs is motivated by the need to monitor and detect the breakdown of SF$_6$, a commonly used gas in electrical equipment such as switchgear and circuit breakers. SF$_6$ is a powerful insulator and arc quencher, but it is also a potent greenhouse gas with a global warming potential of 23,500 times that of CO$_2$. SF$_6$ decomposition products can indicate the presence of partial discharges, corona discharges, or other electrical faults in the equipment, which can lead to equipment failure if left unaddressed. Sensing $SF_6$ decomposition products on GQDs can provide a sensitive and selective method for detecting these breakdowns, and thus, enabling early maintenance and preventing equipment failure. In this current work, we computationally study the sensing properties of three different GQDs (phenalenyl, triangulene, and extended triangulene) for gases $SO_2$, $SOF_2$, $SO_2F_2$ and $SF_6$ using a DFT methodology.

## 2. Computational Methodology

As mentioned earlier, we employed the Gaussian16 suit of programs [49] to perform all the DFT calculations presented in this work. In the first step, we optimized the geometries of −$CH_3$, −$COCH_3$ and −$NH_2$ functionalized phenalenyl, triangulene and extended triangulene. In the simulation process the charge is neutral for all the considered structures along with the molecules adsorbated. The multiplicity of doublet, triplet and quartet is utilized for phenalenyl, triangulene and extended-triangulene, respectively, due to presence of one, two and three unpaired electrons. To address the significant role played by van der Waals (vdW) interactions in systems involving adsorption, we utilized the ωB97XD functional. This functional incorporates London dispersion corrections that account for the long-range interactions that arise during the adsorption process. The ωB97XD functional includes Grimme's van der Waals (vdW) correction term of –$C_6/R^6$, also known as the GD2 dispersion model. The mathematical expression for the dispersion correction term is as follows:

$$E_{disp}^{D2} = -S_6 \sum_{i>j}^{N_{atoms}} \frac{C_6^{ij}}{(R_{ij})^6} f_{damp}(R_{ij}) \quad (1)$$

Here, $f_{damp}(R_{ij}) = \frac{1}{1+ a(R_{ij}/R_r)^{-12}}$ represents the damping function. $R_r$ is the sum of vdW radii of the atomic pair $ij$, and the parameter, $a$, governs the power of dispersion corrections. The number of atoms in the system is denoted by $N_{atoms}$ in Eqn. (1). Additionally, $C_6^{ij}$ represents the dispersion coefficient for the atom pair $ij$, while $R_{ij}$ denotes the interatomic distance between them. The value of the fitting parameter $S_6$, used in the damping function to account for the correlation of this additive dispersion term, was included in the ωB97XD functional, and set to 1.0 [50] In addition, we utilized the hybrid functional B3LYP [51-52], which is a combination of the Becke exchange functional (Becke three parameter) with the Hartree-Fock exchange term. The hybrid B3LYP functional incorporates a non-local correlation functional from LYP (Lee, Yang and Parr) and a local correlation functional of the form VWN (Vosko, Wilk, and Nusair). We employed a 6-31G (D) split-valence basis set, which consists of six Gaussian functions for describing inner-shell orbitals and a split-valence set of four Gaussians for the valence orbitals, with subsets of 3 and 1. The gases to be adsorbed were initially placed parallel to the surfaces of the functionalized GQDs, and the whole systems (GQD+adsorbed gas) were then permitted to relax until the gradient forces achieved the predetermined threshold of 0.00045 Hartree. The optimized structure and the molecular orbitals were visualized with GaussView (version 6). The adsorption energy ($E_{ad}$) of $SO_2$, $SOF_2$, $SO_2F_2$ and $SF_6$ molecules on the three functionalized GQDs was computed using the equation [53-54]:

$$E_{ad} = E_{g+GQDs} - (E_g + E_{GQDs}) \qquad (2)$$

where $\boldsymbol{E_{g+GQDs}}$ is the optimized total energy for the gas molecule $\boldsymbol{g}$ = $SO_2$, $SOF_2$, $SO_2F_2$ and $SF_6$ adsorbed over $\boldsymbol{GQDs}$ = phenalenyl, triangulene and extended triangulene, $\boldsymbol{E_g}$ is the optimized total energy of individual gas molecule $\boldsymbol{g}$ and $\boldsymbol{E_{GQDs}}$ is the optimized energy of pristine GQDs (phenalenyl, and triangulene and extended triangulene). Using the definition, negative value of $E_{ad}$ value shows a stable adsorption complex on the GQD. To further check the effect of higher basis set, we have evaluated the $E_{ad}$ using def2-TZVPP triple zeta basis set and the discussion is presented in supplementary material. The formation energy $E_f$ are calculated using equation:

$$E_f = \frac{1}{N}[E_{functionalized\ GQDs} - nE_c - nE_H - nE_{O/N}] \qquad (3)$$

where $E_{functionalized\ GQDs}$ is the total energy of the functionalized phenalenyl, triangulene and extended triangulene system, $E_C, E_H$ and $E_{O/N}$ is the total energy of individual carbon, hydrogen, oxygen or nitrogen. N represents total number of atoms in the system while, $n$ represents individual number of atom respectively. We observed negative formation energy for all systems, conforming thermodynamical stability and also suggest finite possibility of these materials to be synthesize experimentally. The results obtained using ωB97XD functional are presented in the main text, while those corresponding to B3LYP are presented in the supplementary material.

## 3. Results and Discussion

In this study, we focus on three triangular GQDs with unique topology (i) phenalenyl, which have three fused benzene rings and a total spin quantum number of 1/2, (ii) triangulene which has six fused benzene rings and a total spin quantum number of, and (iii) extended triangulene, which has ten fused benzene rings and a total spin quantum number of 3/2. These GQDs have unusual topologies that prevent the formation of an aromatic structure without the presence of unpaired electrons, resulting in high-spin ground states. Additionally, to investigate the effect of different functional groups on the sensing of gases $SO_2$, $SOF_2$, $SO_2F_2$ and $SF_6$ gases, we functionalize phenalenyl, triangulene and extended triangulene with three different groups: methyl (−$CH_3$−), ketone ($COCH_3$) and amino ($NH_2$). These functional groups are chosen based on previous experimental and theoretical studies [23,48,55]. For simplicity, phenalenyl with functional group is denoted as Phe+$CH_3$, Phe+$COCH_3$ and Phe+$NH_2$, triangulene with functional group is denoted as Tri+$CH_3$, Tri+$COCH_3$ and Tri+$NH_2$, and the extended triangulene as Ex-Tri+$CH_3$, Ex-Tri+$COCH_3$ and Ex-Tri+$NH_2$. Our results were obtained using the ωB97XD functional, which includes long-range corrections for a more accurate description of adsorption properties. The results of the B3LYP functional can be found in the supplementary material.

## 3.1 Structural Properties

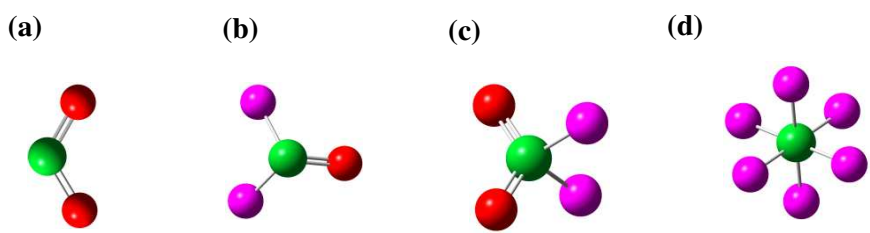

Figure 1: Optimized structures of $SO_2$, $SOF_2$, $SO_2F_2$, and $SF_6$ molecules. The green, red and purple balls represent sulfur, oxygen and fluorine atoms respectively.

Before examining the adsorption process, the structures of the four gas molecules and the three GQDs functionalized with various groups were optimized individually to obtain their most stable configurations. First, we will discuss the optimized structures of the individual gas molecules $SO_2$, $SOF_2$, $SO_2F_2$ and $SF_6$ (see Fig. 1(a-d)). In the case of the $SO_2$ molecule, the bond length of S-O bond is 1.469 Å with bond angle of 119.127°. In $SOF_2$, sulfur is the centre atom bonding with both fluorine atoms and oxygen atom, comprising $sp^2$ hybridization. The S-O and S-F bond lengths are 1.445 Å and 1.627 Å respectively. The F-S-O bond angle is 107.07°, while F-S-F bond angle is 92.652°. In the case of $SO_2F_2$, sulfur is the centre atom bonding with both fluorine atoms and oxygen atoms leading to $sp^3$ hybridization. The S-O and S-F bond lengths are 1.434 Å and 1.586 Å, respectively. In the case of $SF_6$, sulfur is the central atom bonding with six fluorine atoms comprising F-S bond length of 1.6 Å and F-S-F bond angles 90°. These results are in good agreement with the previous reports [56-57].

Next, we will discuss the optimized structures of the functionalized GQDs both with and without the adsorbed molecules, followed by their electronic properties. For the adsorption process, initially the molecules are placed parallel to the GQDs so as to provide them with more access to the surface areas.

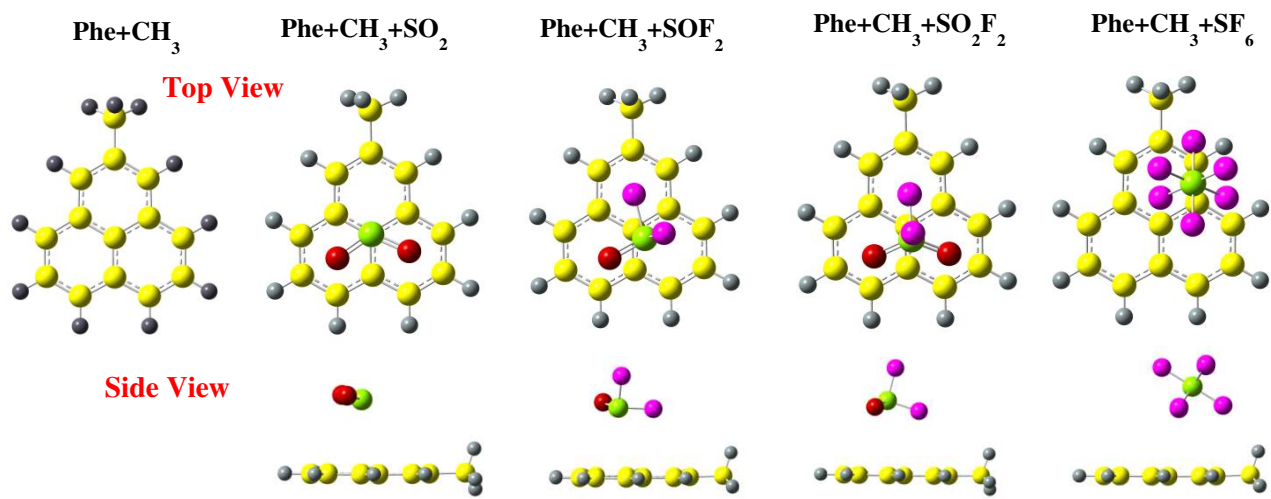

**Figure 2:** Optimized structures of $SO_2$, $SOF_2$, $SO_2F_2$, and $SF_6$ over $CH_3$ edge-functionalized phenalenyl calculated with wB97XD functional.

### 3.1.1 Molecule adsorption on phenalenyl

In Figure 2, we present the optimized structures of isolated Phe+$CH_3$, and the structures with various gas molecules adsorbed on it. For the $SO_2$ molecule, it is located slightly towards the centre of phenalenyl with its oxygen atoms over both hexagonal rings with S-C bond distance of 3.1 Å. The $SO_2$ structure does not present any noticeable modification in the bond lengths, however, the O–S–O bond angle has decreased from 119.127° to 118.42°. In the case of $SOF_2$ adsorption on Phe+$CH_3$, the S-C adsorption distance changes from 2 Å to 3.2 Å with no modification in S–O and S–F bond lengths. The bond angle of F–S–O decreases from 107.07° to 106.50° and the bond angle of F–S–F changes from 92.652° to 92.126°. When $SO_2F_2$ is adsorbed on Phe+$CH_3$, the S-C adsorption distance becomes 3.5 Å with minor to no change in bond lengths and bond angles of the gas molecule. The $SO_2F_2$ molecule is slightly shifted with fluorine atom tilting towards the carbon atoms. In the case of $SF_6$ adsorption on Phe+$CH_3$, the S-C adsorption distance is the highest with 4.47 Å, as compared to all other cases considered here. The bond lengths and bond angles of $SF_6$ depict no change in its structure as molecule is shifted far away and towards the edge of Phe+$CH_3$.

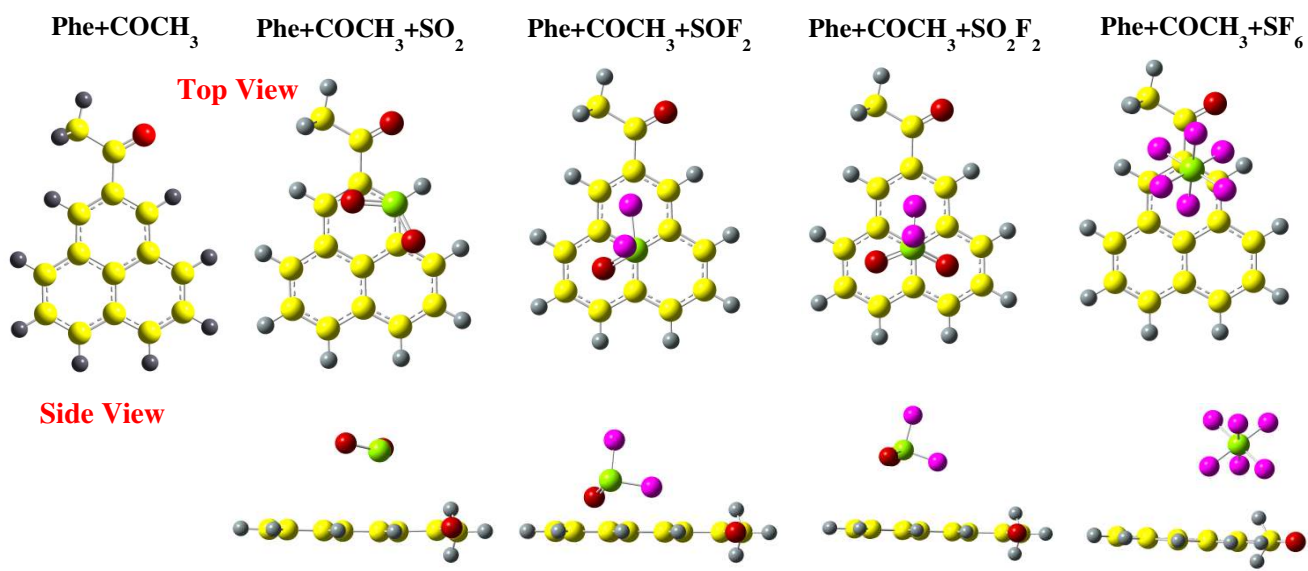

**Figure 3:** Optimized structures of $SO_2$, $SOF_2$, $SO_2F_2$, and $SF_6$ over $COCH_3$ edge-functionalized calculated with wB97XD functional.

In Figure 3, we present the optimized structure of Phe+$COCH_3$ with various gas molecules ($SO_2$, $SOF_2$, $SO_2F_2$ and $SF_6$) adsorbed on it. Similar to the case of $SO_2$ adsorption on Phe+$CH_3$, the molecule is located slightly towards the edge, closer to the oxygen atom of functional group. The O–S–O bond angle has decreased from 119.127° to 118.35° (lower than Phe+$CH_3$). In the case of $SOF_2$ adsorption on Phe+$COCH_3$, the molecule is slightly shifted with S-C distance 3.2 Å, similar to the case of $SOF_2$ over Phe+$CH_3$. The bond angle of F–S–O decrease from 107.07° to 106.62°, and bond angle of F–S–F changes from 92.652° to 92.28°. In the case of $SO_2F_2$ molecule over Phe+$COCH_3$, the adsorbed molecule has not been shifted, but the adsorption distance changes to 3.5 Å with fluorine atom facing the GQD. Similar to $SF_6$ adsorption over Phe+$CH_3$, $SF_6$ is attracted towards the functional group making the initial S-C distance of 4.7 Å with no noticeable alteration in bond lengths and bond angles.

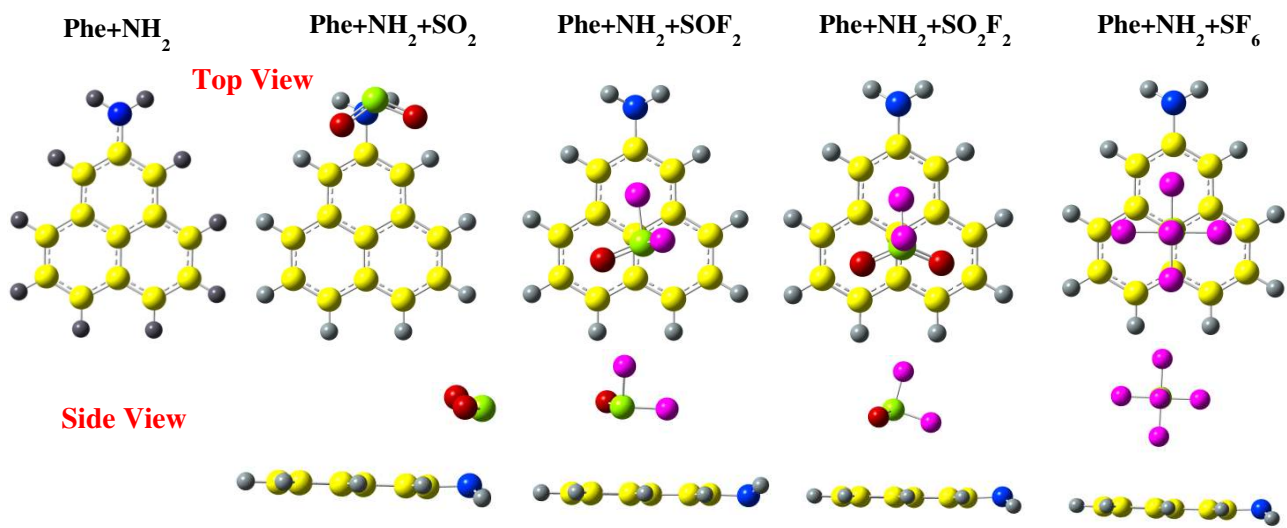

**Figure 4:** Optimized structures of $SO_2$, $SOF_2$, $SO_2F_2$, and $SF_6$ over $NH_2$ edge-functionalized phenalenyl calculated with wB97XD functional.

Figure 4 shows the optimized structures of gas molecules on Phe+$NH_2$. In the case of $SO_2$ adsorption, the molecule positions itself slightly towards the edge of the GQD, with the sulfur atom facing the hydrogen atom of –$NH_2$ functional group. The adsorption distance between the sulfur and nitrogen atom is 2.62 Å and the O–S–O bond angle has decreased from 119.127° to 117.88°. For the $SOF_2$ molecule, the adsorption is similar to that on both –$CH_3$ and –$COCH_3$ group, but on the opposite side of the Phe+$NH_2$ GQD. The bond angle of F–S–O decreases from 107.07° to 106.5° and the bond angle of F–S–F changes from 92.652° to 92.09°. The $SO_2F_2$ molecule adsorbed on Phe+$NH_2$ presents similar results as on Phe+$CH_3$. In contrast, $SF_6$ adsorption on Phe+$NH_2$ results in contrasting structural outcomes as compared to the other two functional groups, –$CH_3$ and –$COCH_3$. The molecule is repelled away from the GQD, with the S-F distance increasing from 2 Å to 2.9 Å.

### 3.1.2 Molecule adsorption on triangulene

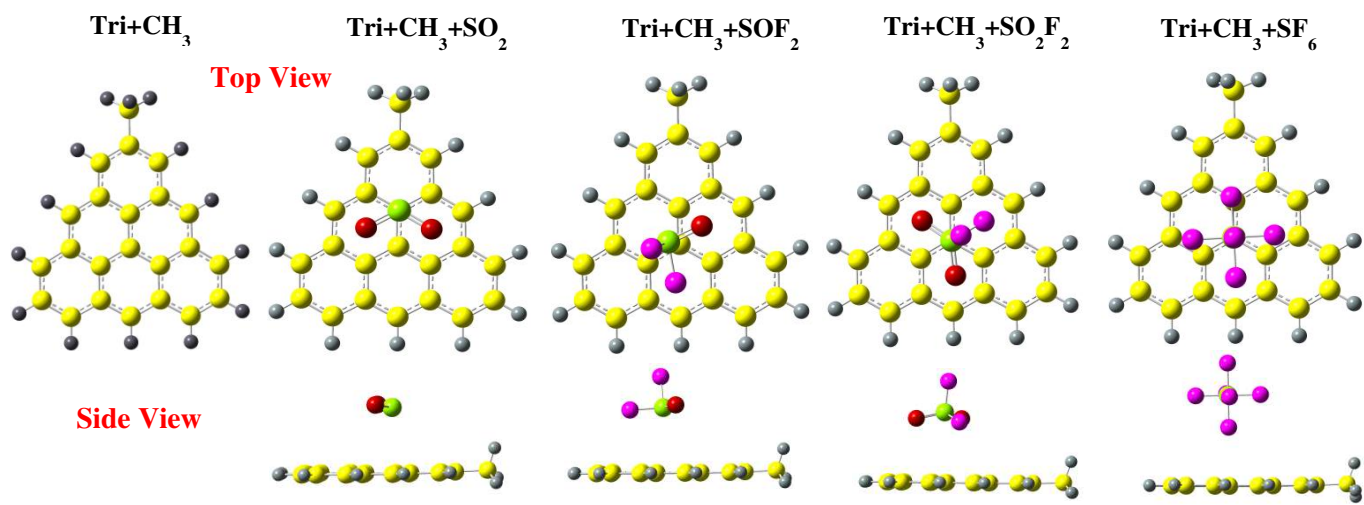

**Figure 5:** Optimized structures of $SO_2$, $SOF_2$, $SO_2F_2$, and $SF_6$ over $CH_3$ edge-functionalized triangulene calculated with wB97XD

Figure 5 presents the optimized structure of Tri+CH$_3$ with and without gas molecules. The adsorption of $SO_2$ on Tri+CH$_3$ is similar to that on Phe+CH$_3$, with a similar adsorption distance and position after optimization. The bond angles are also analogous to those of the $SO_2$ molecule adsorbed on Phe+CH$_3$. The $SOF_2$ molecule realigns itself to the opposite edge of functional group –CH$_3$ with no changes in the S–O and S–F bond lengths. The bond angle of F–S–O decreases from 107.07° to 106.64° and bond angle of F–S–F changes from 92.652° to 92.23°. Similarly, to $SOF_2$, $SO_2F_2$ is also repelled from the –CH$_3$ functional group to the edges of Tri+CH$_3$. In contrast to the $SF_6$ adsorption on Phe+CH$_3$, $SF_6$ on Tri+CH$_3$ is moved slightly away from the functional group with the adsorption distance of 2.9 Å. However, similar to $SF_6$ on Phe+CH$_3$, there are no obvious changes in bond lengths or bond angles.

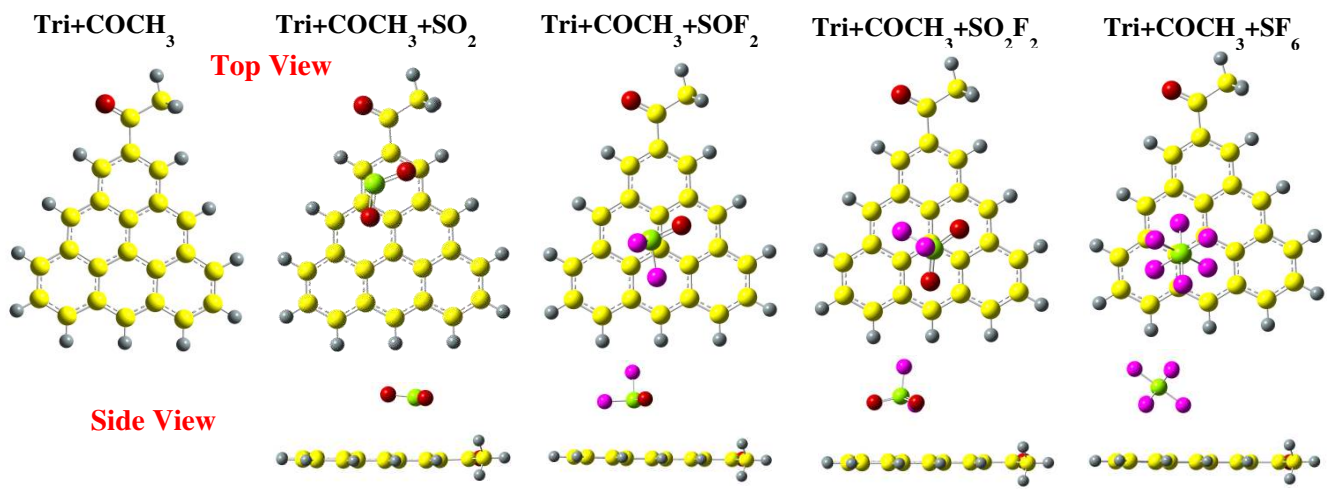

**Figure 6:** Optimized structures of $SO_2$, $SOF_2$, $SO_2F_2$, and $SF_6$ over $COCH_3$ edge-functionalized triangulene calculated with wB97XD.

Figure 6 presents the optimized structure of Tri+$COCH_3$ along with gases. Unlike the adsorption of $SO_2$ over Phe+$COCH_3$, the molecule positions itself towards the methyl group of Tri+$COCH_3$. The O–S–O bond angle has decreased from 119.127° to 118.42°. The structure of $SOF_2$ and $SO_2F_2$ are similar to their adsorption on Tri+$CH_3$. In contrast to $SF_6$ adsorption over Phe+$COCH_3$, $SF_6$ is repelled away from the system, with a final adsorption distance of 3.9 Å.

Figure 7 presents the optimized structures of $SO_2$, $SOF_2$, $SO_2F_2$ and $SF_6$ adsorbed over Tri+$NH_2$. The sulfur atom of $SO_2$ molecule is attracted towards the nitrogen atom of the –$NH_2$ group in Tri+$NH_2$ with an adsorption distance of 2.62 Å. The O–S–O bond angle has decreased from 119.127° to 117.87°. The structural properties of $SOF_2$ are similar to its adsorption on Tri+$COCH_3$. Additionally, the structure of $SO_2F_2$ molecule when adsorbed on Tri+$NH_2$ is similar to its adsorption on Tri+$COCH_3$. The adsorption of $SF_6$ on Tri+$NH_2$ is similar to its adsorption on Phe+$NH_2$, with the $SF_6$ molecule slightly attracted towards the edge of Tri+NH2, with an adsorption distance of 4.6 Å.

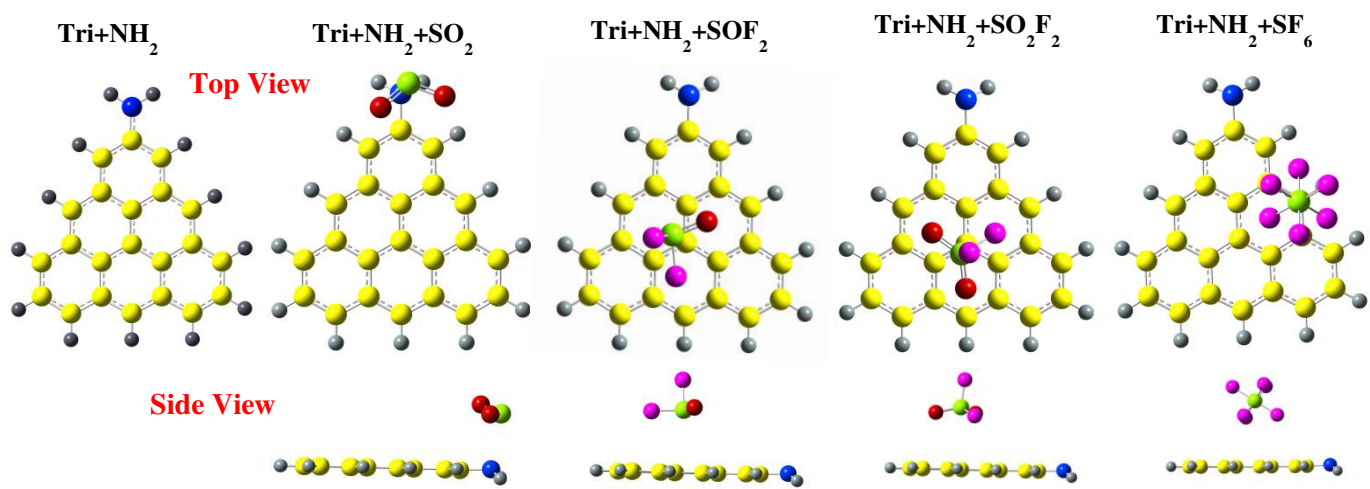

Figure 7: Optimized structures of $SO_2$, $SOF_2$, $SO_2F_2$, and $SF_6$ over $NH_2$ edge-functionalized triangulene calculated with wB97XD functional.

### 3.1.3 Molecule adsorption on extended triangulene

Figure 8 presents the optimized structures of pristine Ex-Tri+$CH_3$, along with those with the gas molecules adsorbed on it. For $SO_2$, the molecule slightly tilts over the carbon atom, resulting in an S-C bond distance of 3.1 Å, which is similar to the case of Phe+$CH_3$. The $SO_2$ structure does not lead to any noticeable changes in the bond lengths, but the O–S–O bond angle decreases from 119.127˚ to 118.48˚. $SOF_2$ adsorption over Ex-Tri+$CH_3$ is similar to that of $SOF_2$ over Tri+$CH_3$.

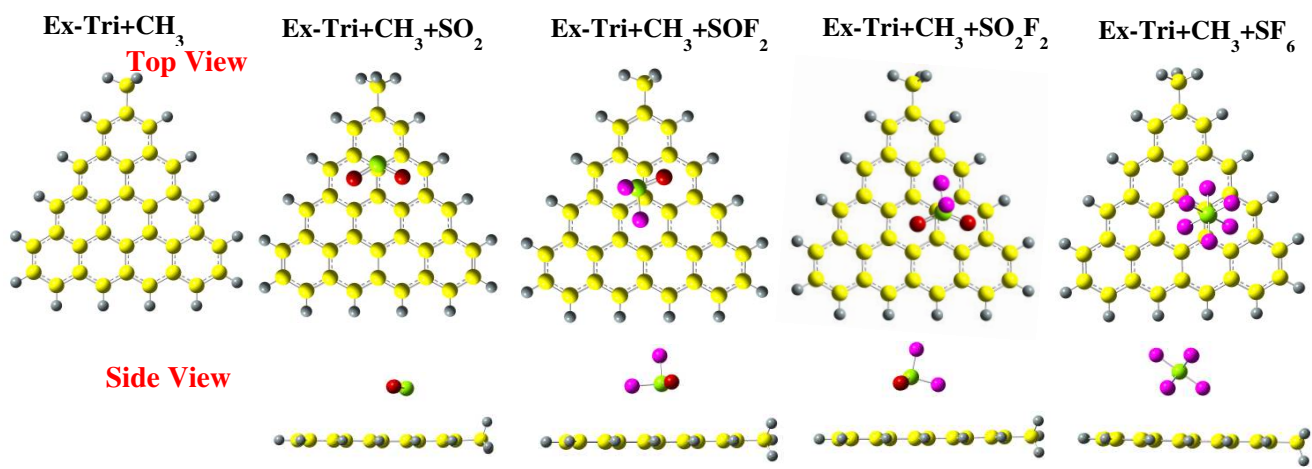

Figure 8: Optimized structures of $SO_2$, $SOF_2$, $SO_2F_2$, and $SF_6$ over $CH_3$ edge-functionalized extended-triangulene calculated with wB97XD functional.

When $SO_2F_2$ is adsorbed over Ex-Tri+$CH_3$, the S-C adsorption distance becomes 3.4 Å with minor to no change in bond lengths and bond angles of the gas molecule. Analogous to Phe+CH$_3$, the SO$_2$F$_2$ molecule slightly shifts towards the edges of Ex-Tri+CH$_3$. The SF$_6$ adsorption on Ex-Tri+CH$_3$ results in a similar structure as in the case of Phe+CH$_3$. Similar adsorption properties are observed when SO$_2$ and SOF$_2$ are adsorbed over Ex-Tri+COCH$_3$ (See Fig. 9). However, both SO$_2$F$_2$ and SF$_6$ are shifted towards the edge of GQD after adsorption, with similar adsorption distances as in case of Ex-Tri+CH$_3$.

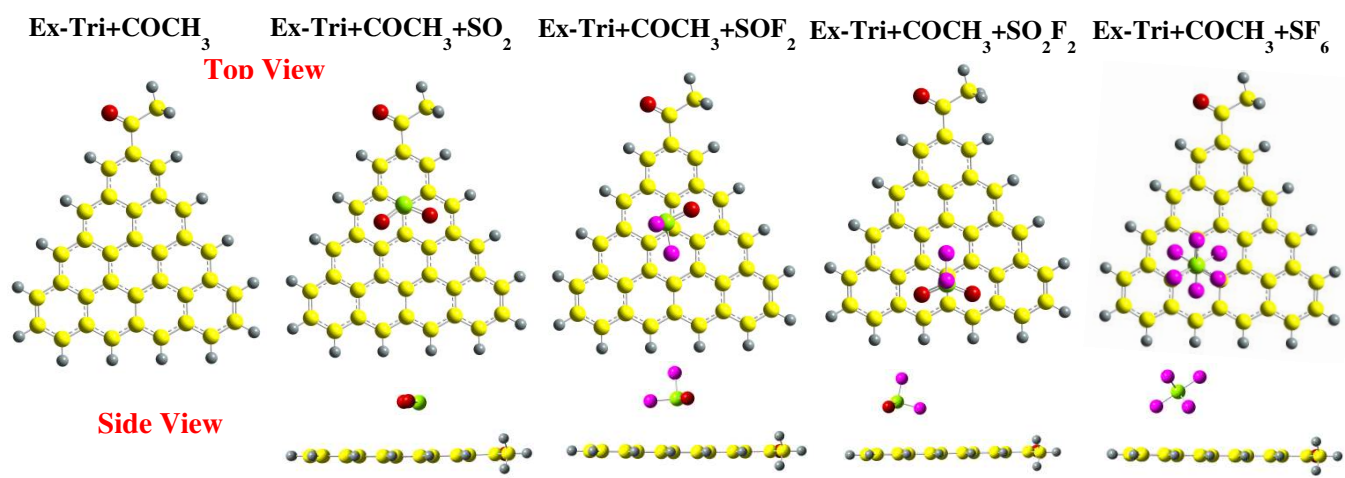

Figure 9: Optimized structures of SO$_2$, SOF$_2$, SO$_2$F$_2$, and SF$_6$ over COCH$_3$ edge-functionalized extended-triangulene calculated with wB97XD.

Fig 10 presents the optimized structures of SO$_2$, SOF$_2$, SO$_2$F$_2$ and SF$_6$ adsorption over Ex-Tri+NH$_2$. SO$_2$ is attracted towards the NH$_2$ functional group with adsorption distance between S and C atom 3.01 Å. The adsorption of SOF$_2$, SO$_2$F$_2$ and SF$_6$ results in similar configurations except for the adsorption distances, which, for the three molecules become 3.19 Å, 3.4 Å, and 3.9 Å, respectively.

To confirm the thermodynamical stability, we have calculated the formation energy ($E_f$) using Eq (3) for functionalized GQDs. The $E_f$ for phenalenyl functionalized with -CH$_3$, -COCH$_3$ and -NH$_2$ are -6.314 eV, -6.957 eV and -6.388 eV, respectively. While in case of $E_f$ for triangulene functionalized with -CH$_3$, -COCH$_3$ and -NH$_2$ are -6.729 eV, -7.153 eV and -6.79 eV, respectively. Further, $E_f$ for extended triangulene functionalized with -CH$_3$, -COCH$_3$ and -NH$_2$ are -7.029 eV, -7.072 eV and -7.090 eV, respectively. The negative $E_f$ for all GQDs confirms their thermodynamical stability and possible fabrication experimentally. Raman calculation is a powerful analytical tool for characterizing the structural stability of materials. The Raman spectrum is sensitive to the vibrational modes of a material and can provide information about the chemical bond structure and symmetry of a material. By monitoring changes in the Raman plot, researchers can determine if a material is undergoing structural changes or if its structure is remaining stable. The Raman plot can be used as a measure of the stability of the structure of a material over time. Raman analysis (presented in Fig S8 of Supplementary material) suggests that all the systems are dynamically stable as there are no imaginary frequencies.

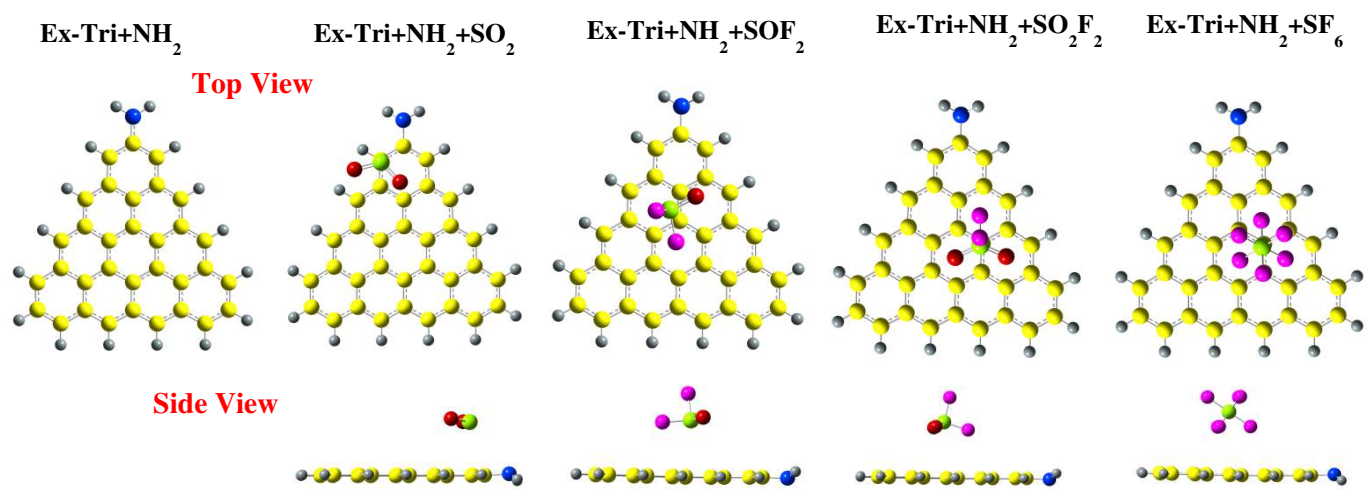

Figure 10: Optimized structures of $SO_2$, $SOF_2$, $SO_2F_2$, and $SF_6$ over $NH_2$ edge-functionalized extended-triangulene calculated with wB97XD functional.

### 3.2 Electronic and Adsorption Properties

Calculations regarding the sensing ability are made using the principle that as the adsorbed gas molecules interact with the GQDs, the electron distribution in the system must change. This electron reorganization should manifest as a measurable change in the ability to conduct an electrical current, known as conductivity. The fundamental parameter that can be calculated using this is the molecular orbits or HOMO-LUMO gap. The electron donating and accepting ability of a system can be defined using the value of HOMO and LUMO energy. These molecular orbitals play vital role in electronic and optical properties, luminescence, photochemical reaction, UV-VIS, quantum chemistry etc. The molecular orbitals help in predicting the most reactive position of the studied system. Further, Mulliken charge transfer is calculated and tabulated in Tables 1-3, confirming the sensitivity of gas molecules $SO_2$, $SOF_2$, $SO_2F_2$ and $SF_6$ over phenalenyl, triangulene and extended triangulene.

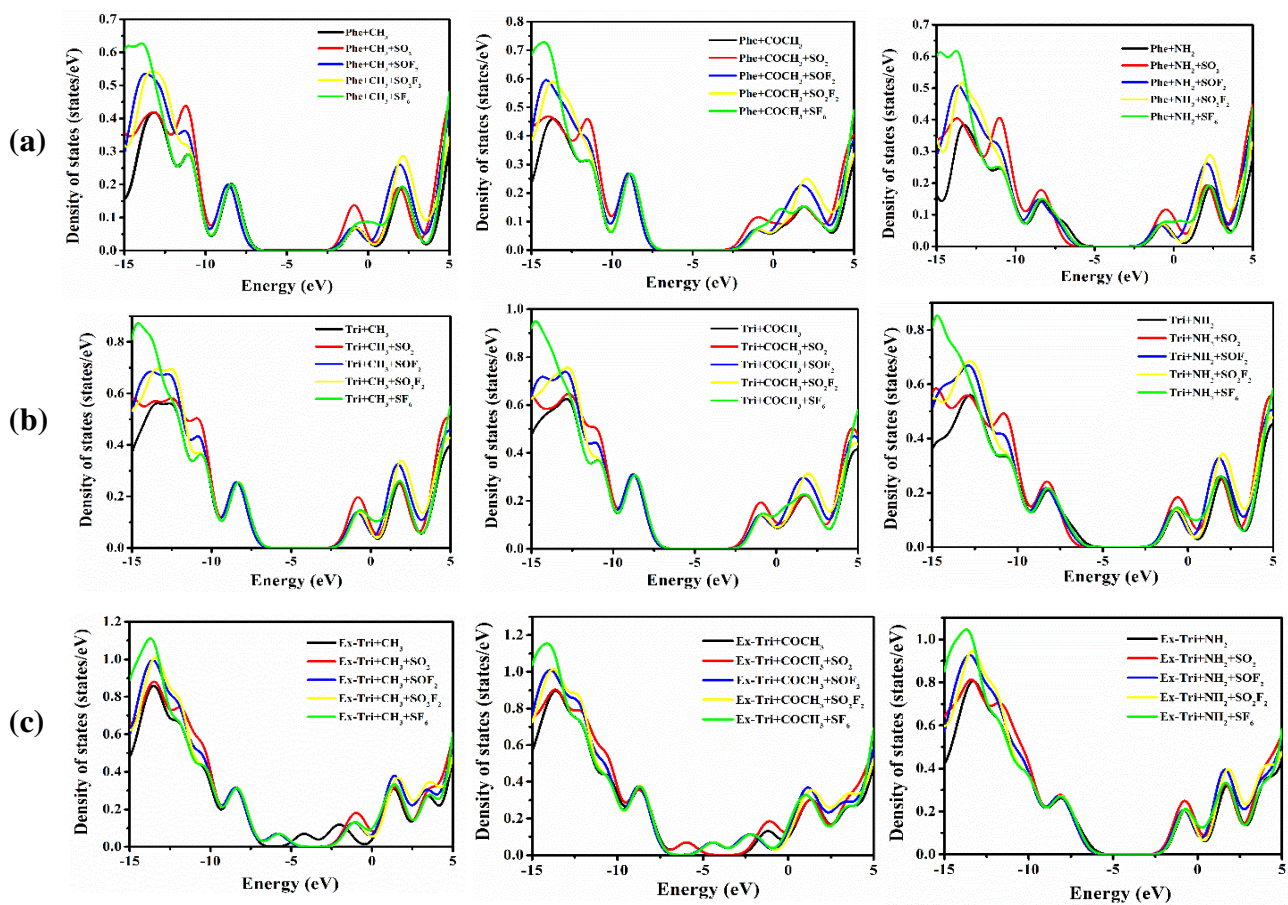

**Figure 11: Density of states of adsorption of $SO_2$, $SOF_2$, $SO_2F_2$, and $SF_6$ with (a) phenalenyl (b) triangulene and (c) extended-triangulene systems with wB97XD functional.**

Figure 11(a-c) depicts total density of states (TDOS) to further analyse the electronic structure from the perspective of size of the GQDs (phenalenyl, triangulene and extended triangulene), functional groups, and adsorbed molecules $SO_2$, $SOF_2$, $SO_2F_2$ and $SF_6$. The total density of states gives the number of states present per energy range. The discrete levels are modified using Gaussian $\frac{1}{\sqrt{2\pi}\alpha}\exp[-\frac{(\varepsilon-\varepsilon_i)^2}{2\alpha^2}]$ as the broadening function. The adsorption of these gas molecules on Phe+$CH_3$ leads to minor changes in highest occupied molecular orbitals (HOMO) and lowest unoccupied molecular orbitals (LUMO). However, the highest

peak of Phe+$CH_3$ located around -13eV in valence band shifts and even splits in case of Phe+$CH_3$+$SF_6$. The HOMO and LUMO in Phe+$CH_3$+$SOF_2$ and Phe+$CH_3$+$SO_2F_2$ depict minor changes attributable to slight alteration in their structural properties. The energy HOMO-LUMO gap(H-L gap) of all considered systems are presented in Table 1.

The DOS plots of the gas molecules adsorbed on Phe+$COCH_3$ show that the HOMO exhibit trivial difference in all cases except Phe+$COCH_3$+$SO_2$. However, the LUMOs are shifted slightly towards higher energies. The molecules adsorbed over Phe+$NH_2$ largely show similar features except that a shift in HOMO is observed for the case of Phe+$NH_2$+$SO_2$. In all three cases of functionalized phenalenyl, $SF_6$ adsorbed systems acquire the highest peak in the valence region. The H-L gap changes significantly in case of Phe+$NH_2$ as compared to others.

Table 1: Energy H-L gap($E_g$), Adsorption Energy ($E_{ad}$) and Recovery time ($\tau$) of structure of $SO_2$, $SOF_2$, $SO_2F_2$, and $SF_6$ adsorption on phenalenyl.

| Structure | Charge (e) | $E_g$ (eV) B3LYP | $E_g$ (eV) ωB97XD | $E_{ad}$ (eV) B3LYP | $E_{ad}$ (eV) ωB97XD | $\tau$ (ps) B3LYP | $\tau$ (ps) ωB97XD |
|---|---|---|---|---|---|---|---|
| Phe+$CH_3$ | - | 3.913 | 7.511 | - | - | - | - |
| Phe+$CH_3$+$SO_2$ | +0.03462 | 3.595 | 7.544 | -0.124 | -0.256 | 121 | 19988 |
| Phe+$CH_3$+$SOF_2$ | +0.02214 | 3.907 | 7.516 | -0.137 | -0.268 | 200 | 31796 |
| Phe+$CH_3$+$SO_2F_2$ | +0.00311 | 3.901 | 7.537 | -0.066 | -0.227 | 12 | 6509 |
| Phe+$CH_3$+$SF_6$ | -0.00998 | 3.696 | 7.502 | -0.074 | -0.184 | 17 | 1233 |
| Phe+$COCH_3$ | - | 4.083 | 7.71 | - | - | - | - |
| Phe+$COCH_3$+$SO_2$ | +0.04581 | 4.108 | 7.40 | -0.26 | -0.241 | 23332 | 11188 |
| Phe+$COCH_3$+$SOF_2$ | +0.01746 | 3.956 | 7.71 | -0.128 | -0.252 | 141 | 17121 |
| Phe+$COCH_3$+$SO_2F_2$ | +0.00183 | 4.041 | 7.724 | -0.064 | -0.223 | 11 | 5576 |
| Phe+$COCH_3$+$SF_6$ | -0.01299 | 4.08 | 7.714 | -0.032 | -0.201 | 3 | 2381 |
| Phe+$NH_2$ | - | 2.847 | 6.38 | - | - | - | - |
| Phe+$NH_2$+$SO_2$ | +0.09231 | 3.359 | 7.01 | -0.37 | -0.429 | 1644093 | 16112404 |
| Phe+$NH_2$+$SOF_2$ | +0.02352 | 2.924 | 6.562 | -0.247 | -0.33 | 14110 | 349899 |
| Phe+$NH_2$+$SO_2F_2$ | +0.00294 | 3.002 | 6.62 | -0.122 | -0.28 | 112 | 50579 |
| Phe+$NH_2$+$SF_6$ | -0.0028 | 2.549 | 6.58 | -0.141 | -0.166 | 233 | 614 |

The middle row in Fig. 11 shows the results of DOS for triangulene functionalized systems. The energy gap changes noticeably only in case of Tri+$CH_3$+$SO_2$, from 7.18 eV to 7.02 eV. The minimum change in the gap is observed in case of Tri+$CH_3$+$SF_6$. Similar changes are also seen also in $SO_2$, $SOF_2$, $SO_2F_2$ and $SF_6$ adsorbed Tri+$COCH_3$ systems. In the case of Tri+$NH_2$, a drastic modification is observed when $SO_2$ is adsorbed, leading to a change in the energy H-L gap from 6.31 eV to 6.85 eV. The minimum H-L gap change is observed in $SOF_2$ adsorbed systems. The energy H-L gap values of Ex-Triangulene are presented in Table 3 with B3LYP and ωB97XD. The largest HOMO-LUMO gap ($E_g$) is found for Ex-Tri+$NH_2$ among all functionalized GQDs. The ωB97XD (long range corrected) predicted energy gaps are considerably larger than the B3LYP predicted gap values. This is mainly due to the inclusion of full Hartree–Fock exchange term in the ωB97XD functional at long distances [58]. It is concluded that for electronic properties such as the HOMO-LUMO gaps, B3LYP shows better results; however, ωB97XD shows better results for studying the adsorption mechanism. The sizes and functionalization of GQDs strongly influence its structural and electronic properties. The energy gap value is a crucial parameter in defining the electrical conductivity (σ) of materials because the energy required to take out an electron from the outer shell to become a free portable charge carrier is proportional to the H-L gap. The link between the $E_g$ and σ of a material can be mathematically represented by the following formula [59]:

$$\sigma \propto e^{\frac{-E_g}{2kT}} \tag{4}$$

where k and T are the Boltzmann's constant and the temperature respectively. The conductivity of a material is inversely proportional to its H-L gap, as shown by equation (4). Therefore, a smaller H-L gap value leads to higher σ at a given temperature T. Consequently, when gas molecules are adsorbed onto the surfaces of graphene quantum dots, a significant decrease in their H-L gap value leads to an increase in their conductivity. The reduction of $E_g$ is observed in case of Phe+$CH_3$+$SF_6$, Phe+$COCH_3$+$SO_2$, Tri+$CH_3$+$SO_2$, Tri+$COCH_3$+$SO_2$, Ex-Tri+$CH_3$+$SO_2$ and Ex-Tri+$COCH_3$+$SO_2$. As a result, the conductivity of the functionalized GQDs increases, providing evidence of the robust interaction between the adsorbed gas molecule and the GQDs. This alteration in the molecular orbitals of the GQDs because of the adsorbed molecules could be identified electronically, which suggests its potential application in sensor technology. The adsorption energies calculated through equation (2) for all optimized systems are given in Table 1, Table 2 and Table 3. The gas molecules get adsorbed on the energetically stable structures of functionalized GQDs, typically at distances in the range 2.46-4.47 Å. These considerably large adsorption distances prohibit the formation of chemical bonds, resulting in physisorption [60-64].

**Table 2: Energy** H-L gap ($E_g$), Adsorption Energy ($E_{ad}$) and Recovery time ($\tau$) of structure of $SO_2$, $SOF_2$, $SO_2F_2$, and $SF_6$ adsorption on triangulene.

| Structure | Charge(e) | $E_g$ (eV) | | $E_{ad}$ (eV) | | $\tau$ | |
|---|---|---|---|---|---|---|---|
| | | B3LYP | ωB97XD | B3LYP | ωB97XD | B3LYP | ωB97XD |
| Tri+$CH_3$ | - | 3.669 | 7.182 | - | - | - | - |
| Tri+$CH_3$+$SO_2$ | +0.03745 | 3.23 | 7.025 | -0.119 | -0.261 | 99 | 24253 |
| Tri+$CH_3$+$SOF_2$ | +0.01667 | 3.62 | 7.192 | -0.111 | -0.282 | 73 | 54644 |
| Tri+$CH_3$+$SO_2F_2$ | +0.00028 | 3.653 | 7.190 | -0.065 | -0.257 | 12 | 20774 |
| Tri+$CH_3$+$SF_6$ | -0.00356 | 3.668 | 7.180 | -0.021 | -0.145 | 2 | 272 |
| Tri+$COCH_3$ | - | 3.734 | 7.267 | - | - | - | - |
| Tri+$COCH_3$+$SO_2$ | +0.03396 | 3.261 | 6.93 | -0.109 | -0.247 | 67 | 14110 |
| Tri+$COCH_3$+$SOF_2$ | +0.0145 | 3.698 | 7.301 | -0.127 | -0.275 | 136 | 41680 |
| Tri+$COCH_3$+$SO_2F_2$ | -0.0048 | 3.72 | 7.288 | -0.061 | -0.261 | 10 | 24253 |
| Tri+$COCH_3$+$SF_6$ | -0.01106 | 3.73 | 7.265 | -0.03 | -0.243 | 3 | 12088 |
| Tri+$NH_2$ | - | 2.808 | 6.312 | - | - | - | - |
| Tri+$NH_2$+$SO_2$ | +0.09266 | 3.275 | 6.853 | -0.365 | -0.427 | 1354983 | 14912345 |
| Tri+$NH_2$+$SOF_2$ | +0.01645 | 2.918 | 6.488 | -0.177 | -0.335 | 940 | 424556 |
| Tri+$NH_2$+$SO_2F_2$ | +0.00027 | 3.017 | 6.504 | -0.162 | -0.389 | 526 | 3428728 |
| Tri+$NH_2$+$SF_6$ | -0.00811 | 2.973 | 6.493 | -0.09 | -0.254 | 32 | 18499 |

The adsorption energy ($E_{ad}$) calculated for $SO_2$, $SOF_2$, $SO_2F_2$ and $SF_6$ over Phe+$CH_3$ is -0.256 eV, -0.268 eV, -0.227 eV and -0.184 eV respectively. The lower adsorption energy in $SO_2F_2$ and $SF_6$ can be attributed to the relatively larger adsorption distances. In Phe+$COCH_3$, the highest adsorption energy is observed for the sensing of $SOF_2$ with the value -0.252 eV. Of all the gas molecules adsorbed on various functionalized GQDs, the largest adsorption energy of -0.429 eV is achieved for the case of $SO_2$ over Phe+$NH_2$. The adsorption energies of functionalized triangulene for various gas molecules are presented in Table 2. For sensing of $SO_2$, Tri+$NH_2$ (-0.427 eV) provides superior adsorption energy as compared to Tri+$CH_3$ (-0.261 eV) and Tri+$COCH_3$ (-0.247 eV). Similarly, Tri+$NH_2$ is energetically better for sensing of $SOF_2$, $SO_2F_2$ and $SF_6$ in comparison with other triangulene functionalizations. In Table 3, we present the adsorption energies of all the functionalized GQDs for various gas molecules. It is obvious from the table that for GQDs functionalized with the groups $CH_3$, $COCH_3$ and $NH_2$, $SOF_2$ shows high adsorption energies -0.285 eV, -0.273 eV, and -0.344 eV, respectively. To understand the electronic interaction between the gas molecules and functionalized GQDs, we plot in Fig. 12 the frontier molecular orbitals.

**Table 3: Energy H-L gap ($E_g$), Adsorption Energy ($E_{ad}$) and Recovery time ($\tau$) of structure of $SO_2$, $SOF_2$, $SO_2F_2$, and $SF_6$ adsorption on extended triangulene.**

| Structure | Charge (e) | $E_g$ (eV) | | $E_{ad}$ (eV) | | $\tau$ | |
|---|---|---|---|---|---|---|---|
| | | B3LYP | ωB97XD | B3LYP | ωB97XD | B3LYP | ωB97XD |
| Ex-Tri+$CH_3$ | - | 1.342 | 4.566 | - | - | - | - |
| Ex-Tri+$CH_3$+$SO_2$ | +0.04094 | 1.329 | 4.432 | -0.180 | -0.261 | 1055 | 24253 |
| Ex-Tri+$CH_3$+$SOF_2$ | +0.01486 | 1.367 | 4.527 | -0.133 | -0.285 | 171 | 61371 |
| Ex-Tri+$CH_3$+$SO_2F_2$ | -0.00064 | 1.429 | 4.555 | -0.102 | -0.260 | 51 | 23332 |
| Ex-Tri+$CH_3$+$SF_6$ | -0.01252 | 1.404 | 4.570 | -0.039 | -0.256 | 4 | 19988 |
| Ex-Tri+$COCH_3$ | - | 1.386 | 4.544 | - | - | - | - |
| Ex-Tri+$COCH_3$+$SO_2$ | +0.0358 | 1.183 | 4.407 | -0.142 | -0.246 | 242 | 13575 |
| Ex-Tri+$COCH_3$+$SOF_2$ | +0.01236 | 1.448 | 4.520 | -0.077 | -0.273 | 19 | 38580 |
| Ex-Tri+$COCH_3$+$SO_2F_2$ | -0.00098 | 1.382 | 4.503 | -0.042 | -0.262 | 5 | 25210 |
| Ex-Tri+$COCH_3$+$SF_6$ | -0.01315 | 1.401 | 4.543 | -0.034 | -0.261 | 3 | 24253 |
| Ex-Tri+$NH_2$ | - | 2.767 | 6.22 | - | - | - | - |
| Ex-Tri+$NH_2$+$SO_2$ | +0.05453 | 2.491 | 6.215 | -0.23 | -0.320 | 7310 | 237660 |
| Ex-Tri+$NH_2$+$SOF_2$ | +0.01644 | 2.873 | 6.395 | -0.219 | -0.344 | 4777 | 601390 |
| Ex-Tri+$NH_2$+$SO_2F_2$ | -0.000093 | 2.968 | 6.405 | -0.161 | -0.324 | 506 | 277423 |
| Ex-Tri+$NH_2$+$SF_6$ | -0.01191 | 2.927 | 6.394 | -0.067 | -0.324 | 13 | 277423 |

It is clear from the figure that $SOF_2$, $SO_2F_2$ and $SF_6$ molecules do not contribute to the formation of either HOMO or LUMO. Instead, these orbitals are completely confined to the surfaces of the functionalized GQDs, confirming the physisorption character of the adsorption processes. Furthermore, the HOMO-LUMO gap of $SOF_2$, $SO_2F_2$ and $SF_6$ over the functionalized GQDs is close to the corresponding pristine values (See Table 1, 2 and 3). In contrast, the behaviour of $SO_2$ on the functionalized GQDs is qualitatively different (as shown in Fig 12). The HOMO (in the case of Phe+$CH_3$, Phe+$COCH_3$, Phe+$NH_2$, Tri+$NH_2$) and LUMO (in the case of Tri+$CH_3$, Tri+$COCH_3$, Ex-Tri+$CH_3$, Ex-Tri+$COCH_3$, Ex-Tri+$NH_2$) are shared indicating strong hybridization.

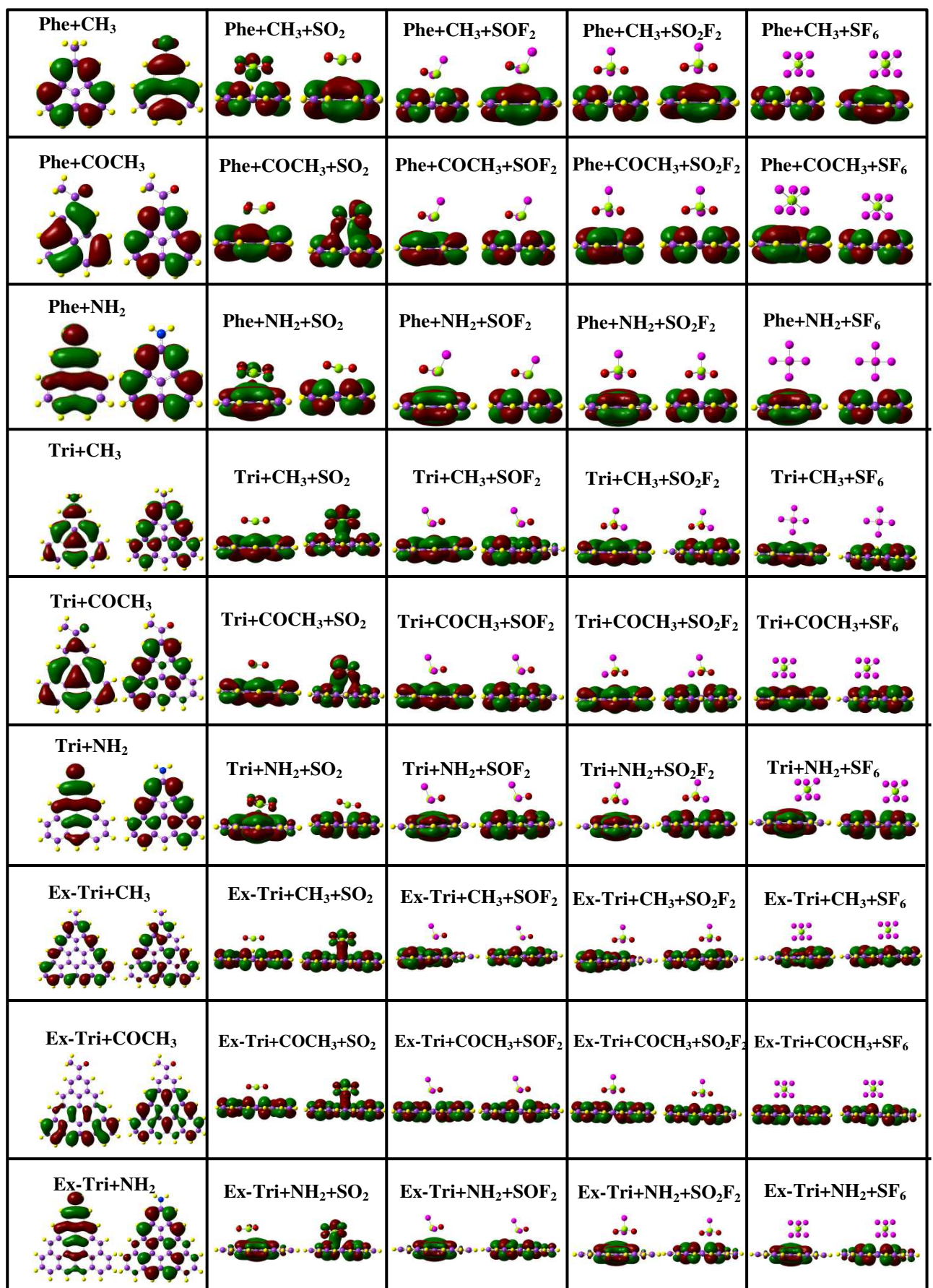

**Figure 12:** HOMO-LUMO plot of adsorption of $SO_2$, $SOF_2$, $SO_2F_2$, and $SF_6$ with phenalenyl, triangulene and extended-triangulene systems with wB97XD functional.

As a result, the HOMO-LUMO gap is altered for these cases. Additionally, non-covalent interaction (NCI) analyses were performed using Multiwfn [64] and VMD [65] packages at the same level of theory to account for H-bonds and non-covalent interactions and are presented in Fig. 13. The strong attraction and repulsion are represented by blue and red region respectively. The van der Waals interaction is shown by the green region. The coloration of the isosurfaces is determined by the values of sign $(\lambda_2)\rho$ (a.u.), ranging from 0.03 to 0.02 a.u. It is observed that strong attractions (blue region) are found in $SO_2$ over $NH_2$ functionalized phenalenyl and triangulene. Steric effect or strong repulsion is found between the hexagonal rings in GQDs, however, van der Waals interaction were observed between the gas molecules and GQDs systems.

## 3.3 Recovery Time

The recovery characteristics of a gas sensor are crucial for determining its usefulness and reusability. The traditional method for evaluating recovery time is by heating the substrate, however, strong chemical reactions or chemisorption between gases and the substrate can result in long recovery times. Theoretical calculations to determine recovery time ($\tau$) can be performed using transition state theory and the Van't-Hoff-Arrhenius expression [66-67], resulting in the expression:

$$\tau = \upsilon^{-1} e^{-E_{ad}/KT} \tag{5}$$

Here, $\upsilon$ is defined as the attempt frequency, K and T denote Boltzmann constant ($\sim 8.318*10^{-3}$ kJ/mol K), and temperature respectively. In the present cases, $10^{12}$ s$^{-1}$ of attempt frequency and 298 K temperature is taken for the recovery time calculation [68-69].

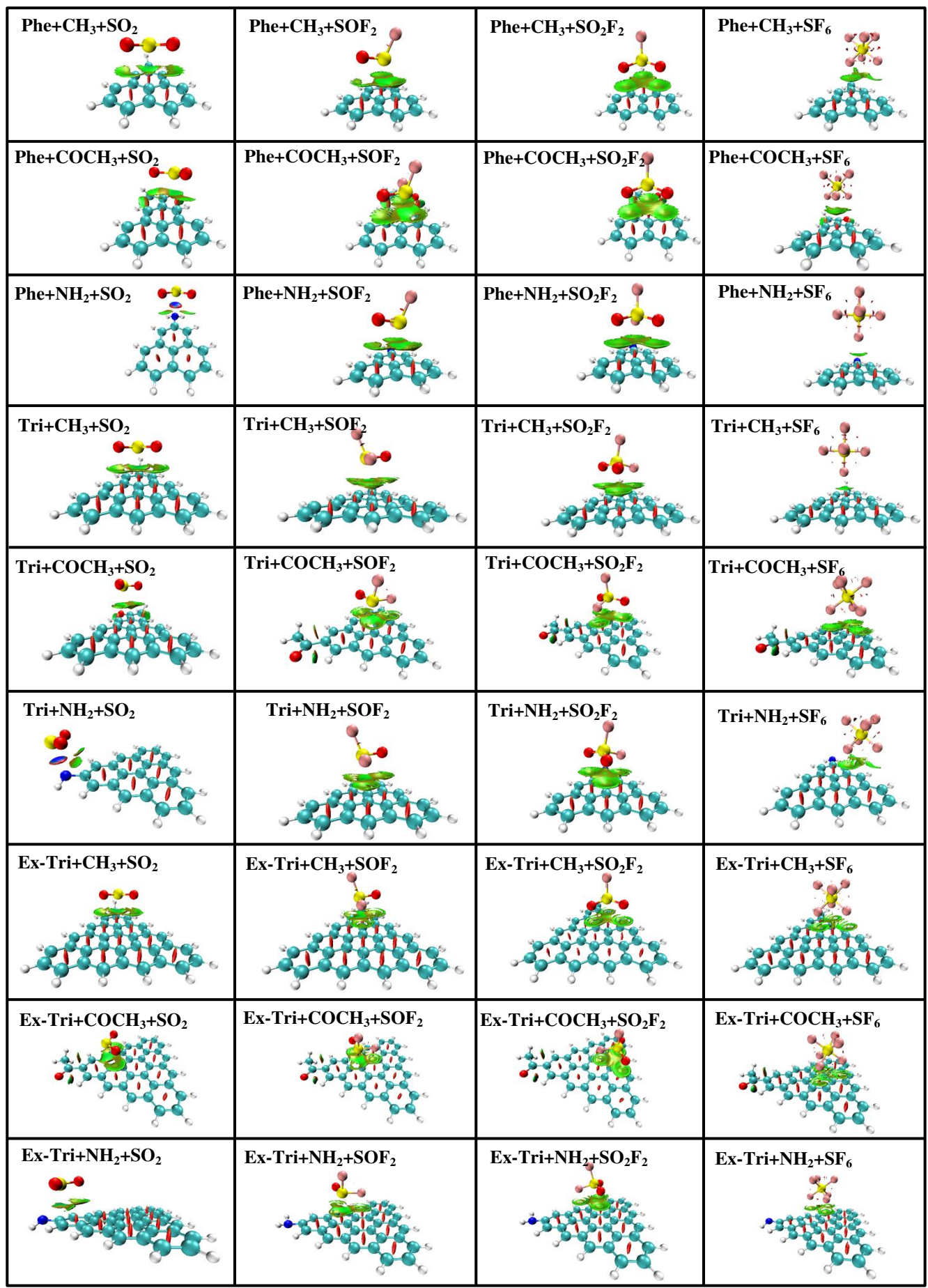

**Figure 13:** NCI plots with wB97XD functional. Blue represents strong attractive interactions, green indicates interactions and red indicates repulsive/steric interactions.

The van der Waals interactions between functionalized GQDs and gas molecules have a significant impact on the recovery time of a gas sensor. This is because van der Waals interactions can strongly influence the adsorption and desorption of gas molecules on the surface of the GQDs. The strength of these interactions can determine the amount of energy required for a gas molecule to adsorb or desorb from the surface, which in turn can affect the recovery time of the sensor. Additionally, the van der Waals interactions can also contribute to a shift in the adsorption energy, which can further impact the performance of the sensor. Therefore, understanding and controlling the van der Waals interactions between functionalized GQDs and gas molecules is crucial for optimizing the performance of gas sensors. Comparing the $-CH_3$, $-COCH_3$ and $-NH_2$ functionalized GQDs according to recovery time, the gas molecules are challenging to desorb from Phe+$NH_2$, Tri+$NH_2$ and Ex-Tri+$NH_2$. The recovery times calculated through equation (5) are given in Tables 1, 2 and 3. In Phe+$CH_3$, $SO_2F_2$ and $SF_6$ have lower recovery times of 6509 ps and 1233 ps, resulting in better desorption property. Similarly for Phe+$COCH_3$, $SO_2F_2$ and $SF_6$ have lower recovery time of 5576 ps and 2381 ps. In summary, both Phe+$CH_3$ and Phe+$COCH_3$ have superior recovery properties and therefore perform better at lower temperatures as compared to Phe+$NH_2$. For sensing $SO_2$ and $SOF_2$, Phe+$NH_2$ and Phe+$COCH_3$ are not very suitable as they require relatively high working temperature for desorption. Similarly to $-NH_2$ functionalized phenalenyl system, Tri+$NH_2$ and Ex-Tri+$NH_2$ also have high recovery times for $SO_2$, $SOF_2$, $SO_2F_2$ and $SF_6$ gases–making them unsuitable for use. In order to put the performance of functionalized GQDs in perspective, we briefly discuss the recovery times of a few other adsorbing materials for the same gas molecules. The α-arsenene with $SO_2$, $SOF_2$ and $SO_2F_2$ shows recovery times of 0.23s, 0.68ms and 0.44μs, however β-arsenene with $SO_2$, $SOF_2$ and $SO_2F_2$ provides 3.65 μs, 0.33 μs and 0.05 μs of recovery times, respectively [70]. The $SO_2$, $SOF_2$ and $SO_2F_2$ have long recovery times of 400 s, 669 μs and 5.9 ns at 298K, respectively for Ni-BNNT system [71]. It is found that in Pt-MoS$_2$ and Au-MoS$_2$, SO$_2$ and SOF$_2$ are difficult to desorb unless increasing the working temperature [72]. These recovery times are higher as compared to GQDs (in ps) in the present study leading to their superiority and high detection limit.

## 4. Conclusions

In this study, we employed density functional theory (DFT) to investigate the adsorption behaviour and gas sensing properties of size and edge-functionalized graphene quantum dots (GQDs), specifically phenalenyl, triangulene, and extended triangulene. We analyzed properties such as adsorption energy, adsorption distance, recovery time, and density of states. Our results show that functionalization of phenalenyl, triangulene, and extended triangulene alters their electronic properties. However, the adsorption of SO$_2$, SOF$_2$, SO$_2$F$_2$, and SF$_6$ gases on Phe+CH$_3$ leads to minor changes in HOMO and LUMO. A significant change is observed when SO$_2$ is adsorbed on -NH$_2$ functionalized triangulene, resulting in a change in the energy gap. Furthermore, -NH$_2$ functionalized phenalenyl, triangulene, and extended triangulene show superior adsorption energies for SO$_2$ sensing compared to other functionalizations. In terms of recovery time, it is challenging to desorb the gases (SO$_2$, SOF$_2$, SO$_2$F$_2$, and SF$_6$) from Phe+NH$_2$, Tri+NH$_2$, and Ex-Tri+NH$_2$ when using -CH$_3$, -COCH$_3$, and -NH$_2$ functionalized phenalenyl and triangulene. In conclusion, this study offers a microscopic understanding of the ultrafast recovery times of GQDs and their potential applications in sensing toxic environmental gases.

**SUPPLEMENTARY MATERIAL**

See supplementary material for the initial configurations for gas molecules adsorption on functionalized phenalenyl and triangulene. Results of B3LYP functional and Raman plots of wB97XD are also presented in supplementary material.

**CRediT authorship contribution statement**

**Vaishali Roondhe**: Data curation, Formal analysis, Writing original draft. **Basant Roondhe**: Methodology, Software. **Sumit Saxena**: Supervision, review & editing. **Rajeev Ahuja**: Supervision, review & editing, **Alok Shukla**: Supervision, Conceptualization, Resources, review & editing.

## Conflicts of interest

There are no conflicts to declare.

## Data availability

Correspondence and requests for data should be addressed to V.R.

## Acknowledgements

Authors VR and BR would like to acknowledge the supported by the Institute Post-Doctoral Fellowship (IPDF) of Indian Institute of Technology Bombay. This article is financially supported by the Swedish Research Council (VR-2016-06014 & VR-2020-04410) and J. Gust. Richert stiftelse, Sweden (2021-00665).

# On Using Non-Kekulé Triangular Graphene Quantum Dots for Scavenging Hazardous Sulfur Hexafluoride Components

**Vaishali Roondhe[1,*], Basant Roondhe[2], Sumit Saxena[2], Rajeev Ahuja[3,4,*] and Alok Shukla[1,*]**

[1]Department of Physics, Indian Institute of Technology Bombay, Mumbai-400076, Maharashtra, India

[2]Department of Metallurgical Engineering and Materials Science, Indian Institute of Technology Bombay, Mumbai-400076, Maharashtra, India

[3]Materials Theory Division, Department of Physics and Astronomy, Uppsala University, Box 516, Uppsala 75120, Sweden

[4]Department of Physics, Indian Institute of Technology Ropar-140001, Punjab, India

## Supplementary Figures

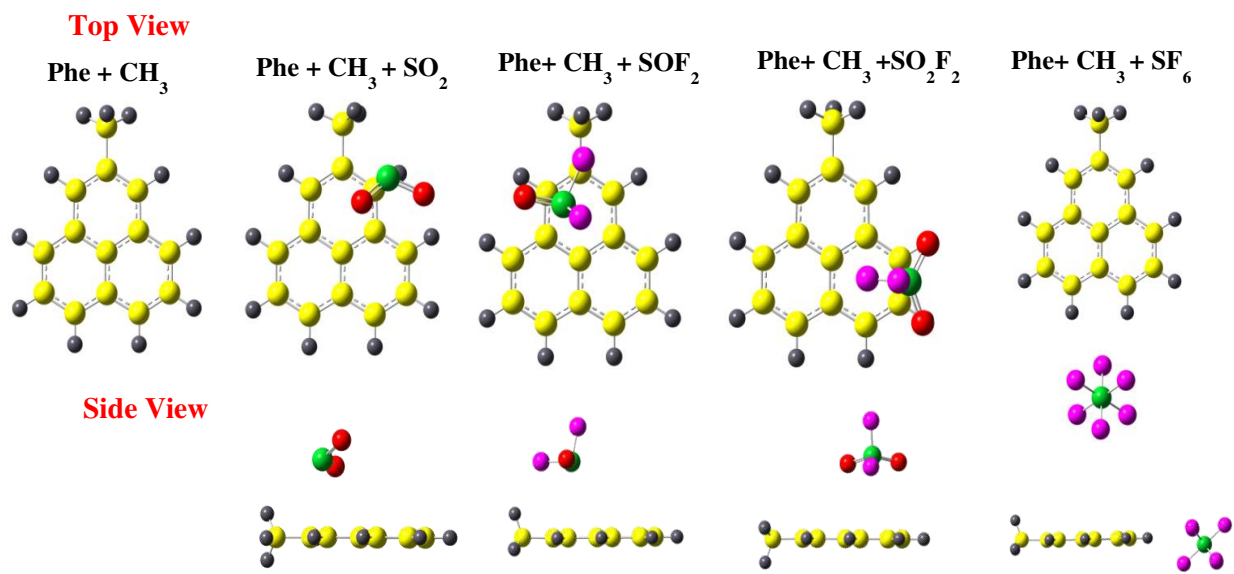

**Figure S1:** Optimized structures of $SO_2$, $SOF_2$, $SO_2F_2$, and $SF_6$ with $CH_3$ edge-functionalized phenalenyl with B3LYP functional. The yellow, grey, green, red and purple balls represent carbon, hydrogen, sulfur, oxygen and fluorine atoms respectively.

---

[*] Corresponding Author:
Email addresses: shukla@phy.iitb.ac.in (A. Shukla), oshivaishali@gmail.com (V. Roondhe), rajeev.ahuja@physics.uu.se (R. Ahuja)

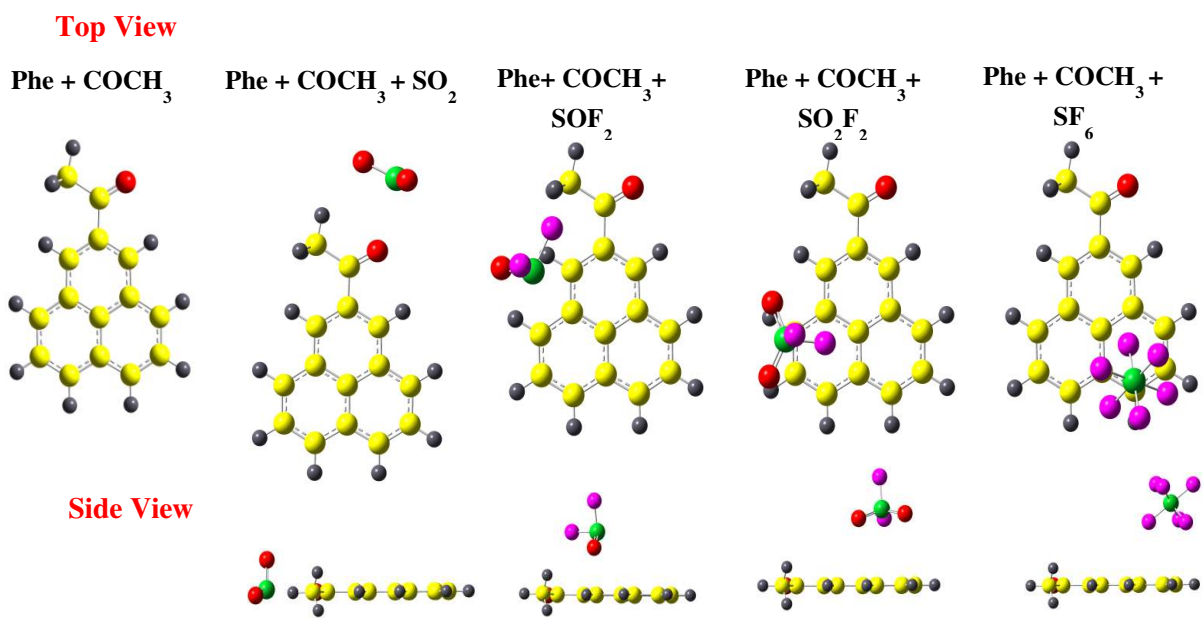

**Figure S2:** Optimized structures of $SO_2$, $SOF_2$, $SO_2F_2$, and $SF_6$ with $COCH_3$ edge-functionalized phenalenyl with B3LYP functional.

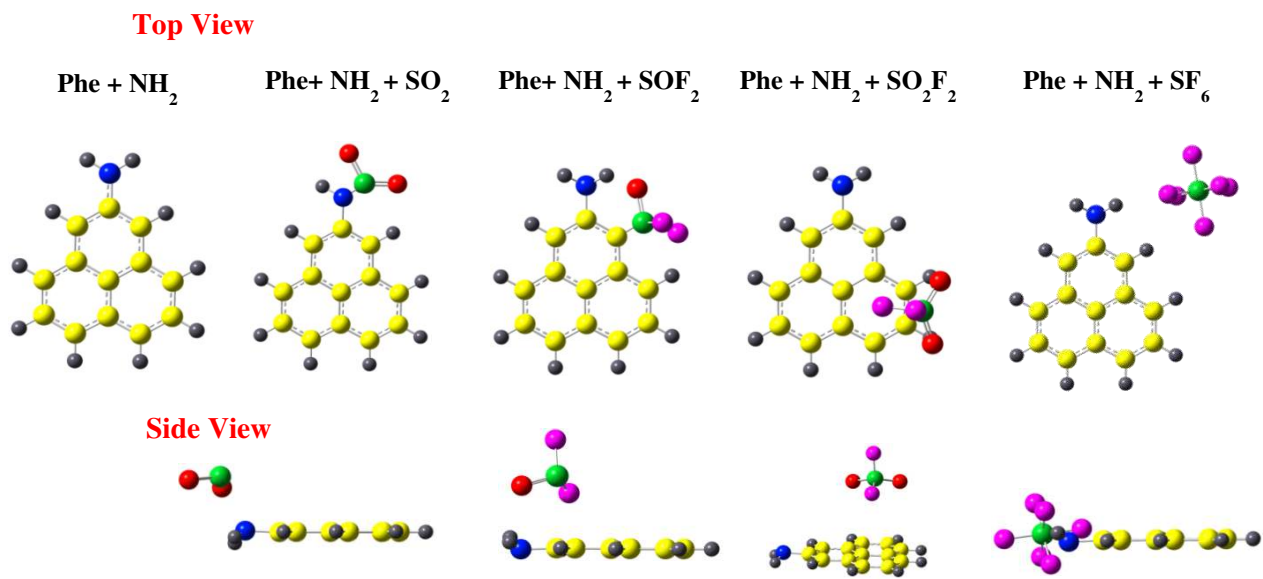

**Figure S3:** Initial structures of $SO_2$, $SOF_2$, $SO_2F_2$, and $SF_6$ with $NH_2$ edge-functionalized phenalenyl with B3LYP functional.

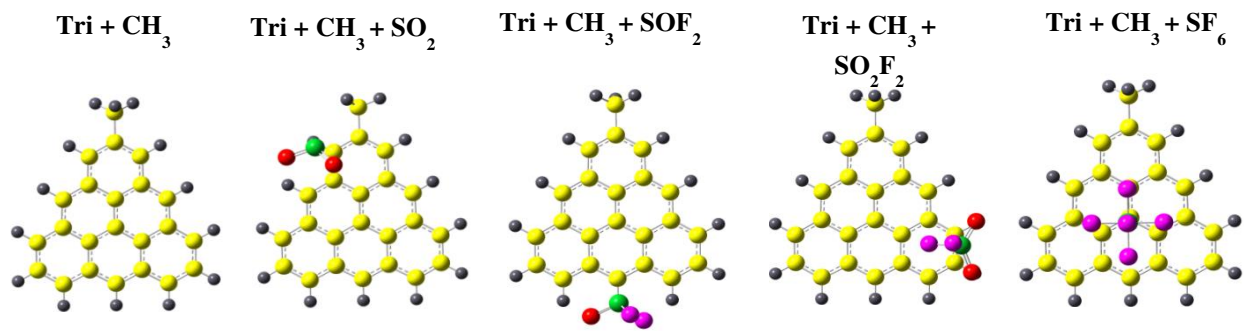

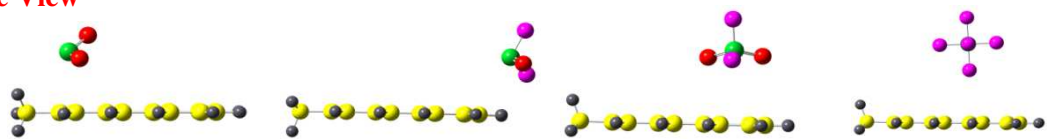

**Figure S4:** Optimized structures of SO$_2$, SOF$_2$, SO$_2$F$_2$, and SF$_6$ with CH$_3$ edge-functionalized triangulene with B3LYP functional.

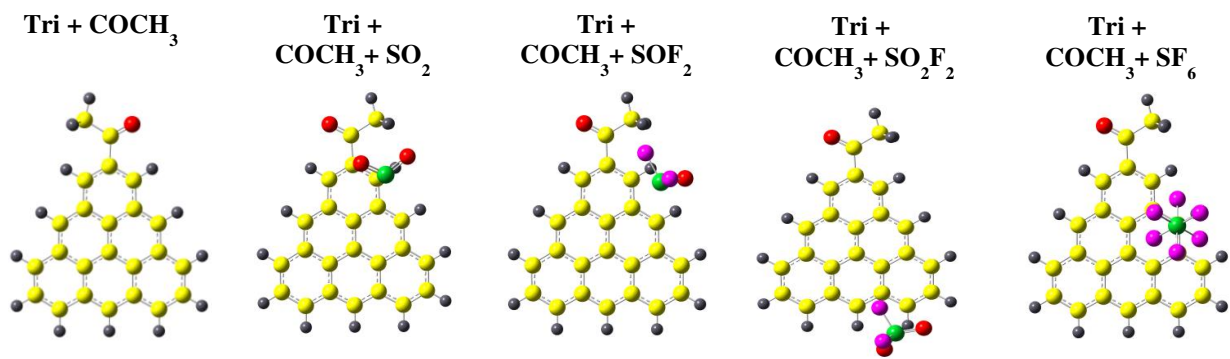

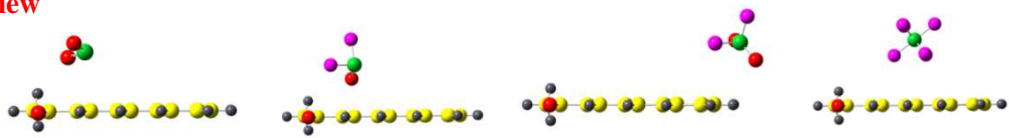

**Figure S5:** Optimized structures of SO$_2$, SOF$_2$, SO$_2$F$_2$, and SF$_6$ with COCH$_3$ edge-functionalized triangulene with B3LYP functional.

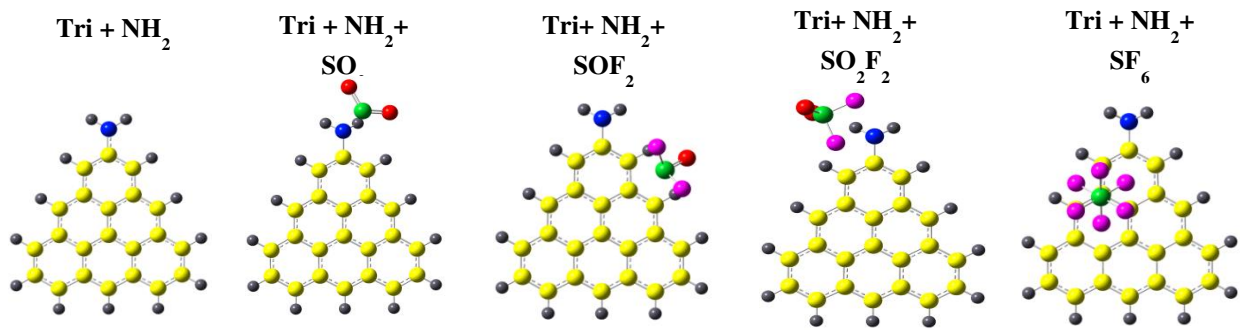
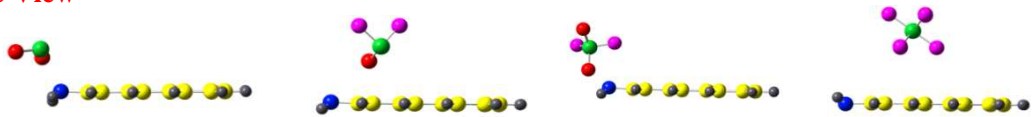

**Figure S6:** Optimized structures of $SO_2$, $SOF_2$, $SO_2F_2$, and $SF_6$ with $NH_2$ edge-functionalized triangulene with B3LYP functional.

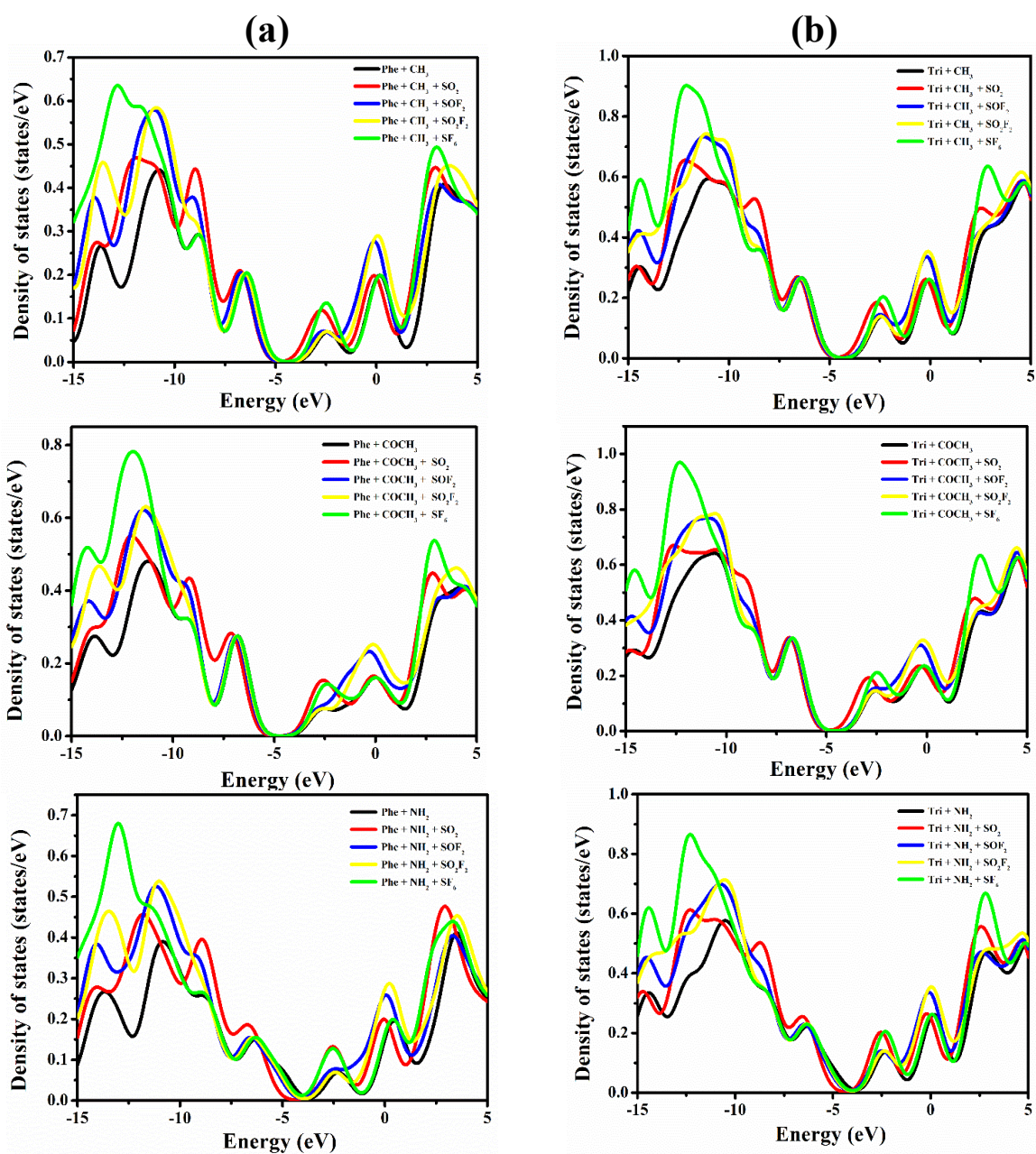

**Figure S7: Density of states of adsorption of $SO_2$, $SOF_2$, $SO_2F_2$, and $SF_6$ with (a) phenalenyl and (b) triangulene systems with B3LYP functional.**

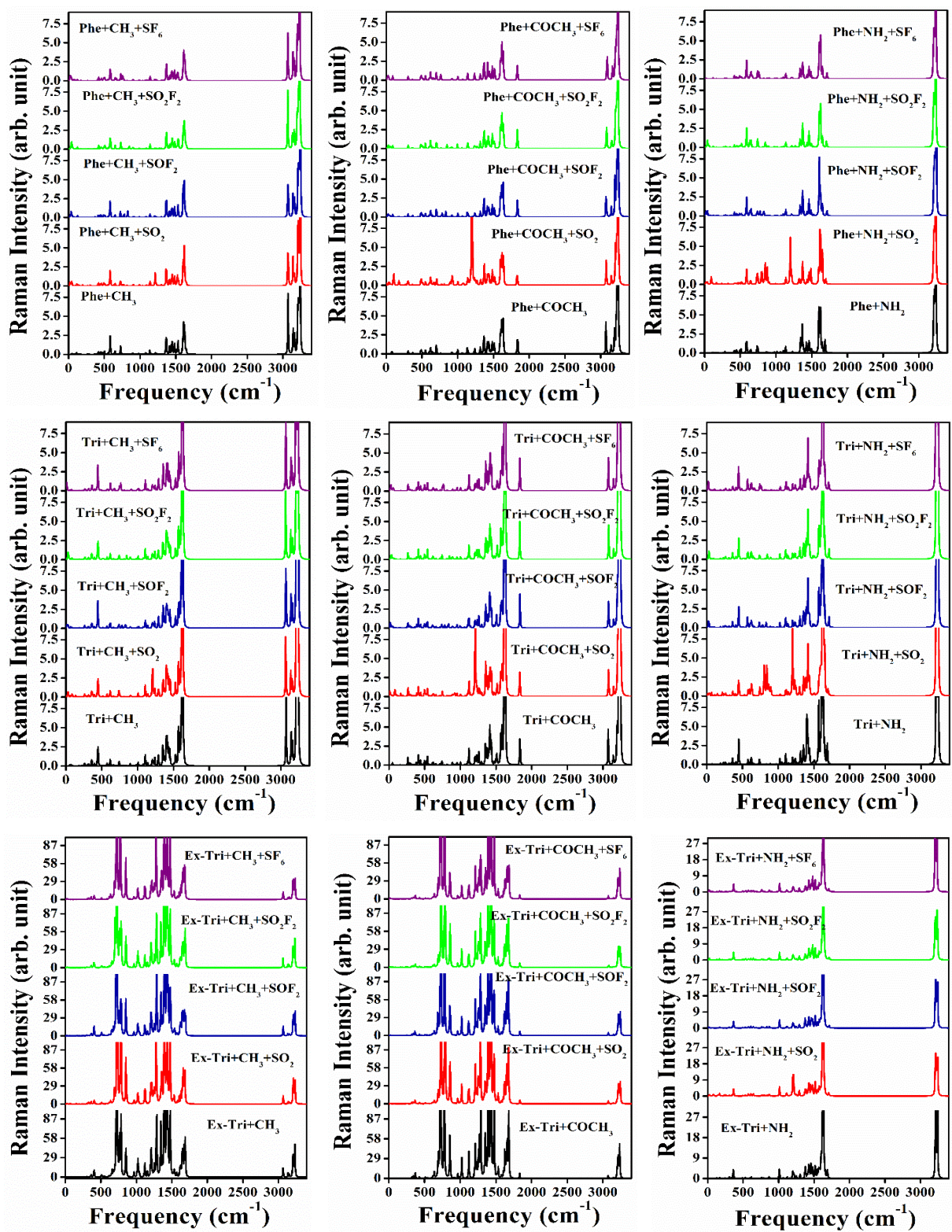

**Figure S8:** Raman plot of adsorption of $SO_2$, $SOF_2$, $SO_2F_2$, and $SF_6$ with phenalenyl (upper row) and triangulene (middle row) and extended triangulene (lower row) systems with wB97XD functional.

**Basis set effect on adsorption energy:-**

We have also done the comparative study on the adsorption energy using higher basis set for all considered GQDs with 6-31G (D) basis set. We found, in our present study, the $E_{ad}$ value with 6-31G (D) basis set is higher as compared to def2-TZVPP triple zeta basis set with single-point calculations. It is possible for the adsorption energy to be lower when using the def2-TZVPP triple zeta basis set as compared to the 6-31G (D) basis set, although the magnitude of the difference will depend on the individual systems being studied. The adsorption energy is a function of many factors, including the electronic structure of the adsorbate and substrate, as well as the adsorption geometry. It is also important to note that the adsorption energy is influenced by many other factors, such as the choice of functional and the accuracy of the geometry optimization. The $E_{ad}$ values obtained using def2-TZVPP triple zeta basis set are tabulated in Table S1.

**Table S1: Adsorption Energy ($E_{ad}$) of structure of $SO_2$, $SOF_2$, $SO_2F_2$, and $SF_6$ adsorption on phenalenyl, triangulene and extended triangulene using def2-TZVPP basis set.**

| Structure | $E_{ad}$ (eV) | Structure | $E_{ad}$ (eV) | Structure | $E_{ad}$ (eV) |
|---|---|---|---|---|---|
| Phe+$CH_3$ | - | Tri+$CH_3$ | - | Ex-Tri+$CH_3$ | - |
| Phe+$CH_3$+$SO_2$ | -0.22 | Tri+$CH_3$+$SO_2$ | -0.23 | Ex-Tri+$CH_3$+$SO_2$ | -0.23 |
| Phe+$CH_3$+$SOF_2$ | -0.21 | Tri+$CH_3$+$SOF_2$ | -0.22 | Ex-Tri+$CH_3$+$SOF_2$ | -0.21 |
| Phe+$CH_3$+$SO_2F_2$ | -0.15 | Tri+$CH_3$+$SO_2F_2$ | -0.18 | Ex-Tri+$CH_3$+$SO_2F_2$ | -0.18 |
| Phe+$CH_3$+$SF_6$ | -0.12 | Tri+$CH_3$+$SF_6$ | -0.11 | Ex-Tri+$CH_3$+$SF_6$ | -0.19 |
| Phe+$COCH_3$ | - | Tri+$COCH_3$ | - | Ex-Tri+$COCH_3$ | - |
| Phe+$COCH_3$+$SO_2$ | -0.20 | Tri+$COCH_3$+$SO_2$ | -0.21 | Ex-Tri+$COCH_3$+$SO_2$ | -0.21 |
| Phe+$COCH_3$+$SOF_2$ | -0.19 | Tri+$COCH_3$+$SOF_2$ | -0.22 | Ex-Tri+$COCH_3$+$SOF_2$ | -0.21 |
| Phe+$COCH_3$+$SO_2F_2$ | -0.15 | Tri+$COCH_3$+$SO_2F_2$ | -0.19 | Ex-Tri+$COCH_3$+$SO_2F_2$ | -0.18 |
| Phe+$COCH_3$+$SF_6$ | -0.13 | Tri+$COCH_3$+$SF_6$ | -0.17 | Ex-Tri+$COCH_3$+$SF_6$ | -0.19 |
| Phe+$NH_2$ | - | Tri+$NH_2$ | - | Ex-Tri+$NH_2$ | - |
| Phe+$NH_2$+$SO_2$ | -0.33 | Tri+$NH_2$+$SO_2$ | -0.33 | Ex-Tri+$NH_2$+$SO_2$ | -0.27 |
| Phe+$NH_2$+$SOF_2$ | -0.25 | Tri+$NH_2$+$SOF_2$ | -0.26 | Ex-Tri+$NH_2$+$SOF_2$ | -0.27 |
| Phe+$NH_2$+$SO_2F_2$ | -0.19 | Tri+$NH_2$+$SO_2F_2$ | -0.22 | Ex-Tri+$NH_2$+$SO_2F_2$ | -0.23 |
| Phe+$NH_2$+$SF_6$ | -0.13 | Tri+$NH_2$+$SF_6$ | -0.17 | Ex-Tri+$NH_2$+$SF_6$ | -0.23 |